\documentclass[journal=ancac3,manuscript=article]{achemso}
\setkeys{acs}{articletitle=true}
\usepackage{graphicx}
\usepackage{epstopdf}
\usepackage{dcolumn}
\usepackage{amsmath}
\usepackage{epsfig}
\usepackage{amssymb}
\usepackage{color}
\usepackage[T1]{fontenc} 

\usepackage{bm}
\usepackage{natbib}
\usepackage[mathlines]{lineno}

\def\didv{\mbox{$dI/d\mbox{V}$}} 
\def\vs{\mbox{V$_{\textrm{s}}$}}
\def\Vs{\mbox{V$_{\textrm{s}}$}}
\def\It{\mbox{I$_{\textrm{t}}$}}
\def\Zoff{\mbox{Z$_{\textrm{off}}$}}

\title
  {Doping of Graphene Nanoribbons via Functional Group Edge Modification}

\author{Eduard Carbonell-Sanrom\`a}
\affiliation{CIC nanoGUNE, Tolosa Hiribidea 76, 20018 Donostia-San Sebastian, Spain}
\altaffiliation{Contributed equally to this work}
\email{e.carbonell@nanogune.eu}

\author{Jeremy Hieulle}
\affiliation{CIC nanoGUNE, Tolosa Hiribidea 76, 20018 Donostia-San Sebastian, Spain}
\altaffiliation{Contributed equally to this work}

\author{Manuel Vilas-Varela}
\affiliation{Centro de Investigaci\'on en Qu\'imica Biol\'oxica e Materiais Moleculares (CIQUS) and Departamento de Qu\'imica Org\'anica, Universidade de Santiago de Compostela, 15782 Santiago de Compostela, Spain}

\author{Pedro Brandimarte}
\affiliation{Centro de F\'isica de Materiales (CSIC-UPV/EHU), 20018 Donostia-San Sebastian, Spain}

\author{Mikel Iraola}
\affiliation{CIC nanoGUNE, Tolosa Hiribidea 76, 20018 Donostia-San Sebastian, Spain}

\author{Ana Barrag\'an}
\affiliation{Centro de F\'isica de Materiales, 20018 Donostia-San Sebastian, Spain}

\author{Jingcheng Li}
\affiliation{Centro de F\'isica de Materiales, 20018 Donostia-San Sebastian, Spain}

\author{Mikel Abadia}
\affiliation{Centro de F\'isica de Materiales, 20018 Donostia-San Sebastian, Spain}

\author{Martina Corso}
\affiliation{CIC nanoGUNE, Tolosa Hiribidea 76, 20018 Donostia-San Sebastian, Spain}
\alsoaffiliation{Centro de F\'isica de Materiales, 20018 Donostia-San Sebastian, Spain}
\alsoaffiliation{Donostia International Physics Center, 20018 Donostia-San Sebastian, 
Spain}

\author{Daniel S\'anchez-Portal}
\affiliation{Centro de F\'isica de Materiales, 20018 Donostia-San Sebastian, Spain}
\alsoaffiliation{Donostia International Physics Center, 20018 Donostia-San Sebastian, 
Spain}

\author{Diego Pe\~na}
\affiliation{Centro de Investigaci\'on en Qu\'imica Biol\'oxica e Materiais Moleculares (CIQUS) and Departamento de Qu\'imica Org\'anica, Universidade de Santiago de Compostela, 15782 Santiago de Compostela, Spain}
\email{diego.pena@usc.es}

\author{Jose Ignacio Pascual}
\affiliation{CIC nanoGUNE, Tolosa Hiribidea 76, 20018 Donostia-San Sebastian, Spain}
\alsoaffiliation{Ikerbasque, Basque Foundation for Science, Bilbao, Spain}
\email{ji.pascual@nanogune.eu}


\begin{document}


\newpage
\begin{abstract}
	
  We report on the on-surface synthesis of 7 armchair graphene nanoribbons (7-AGNRs) substituted with nitrile (CN) functional groups. The CN groups are attached to the GNR backbone by modifying the 7-AGNR precursor.  While many of these groups survive the on-surface synthesis, the reaction process causes the cleavage of some CN from the ribbon backbone and the on-surface cycloisomerization of few nitriles onto pyridine rings. Scanning Tunneling Spectroscopy and Density Functional Theory reveal that CN groups behave as very efficient $n$-dopants, significantly downshifting the bands of the ribbon, and introducing deep impurity levels associated to the nitrogen electron lone pairs.
  
\end{abstract}
\newpage

Graphene nanoribbons (GNRs) have recently arisen as potential materials capable to overcome the absence of an electronic bandgap in graphene, while maintaining many other of their structural and charge mobility properties. Narrow enough armchair graphene nanoribbons (AGNRs) possess a bandgap while zigzag graphene nanoribbons are characterized by edges hosting spin polarized states \cite{fujita_peculiar_1996,wakabayashi_electronic_2010}. On surface synthesis of GNRs has demonstrated to achieve their growth with the atomic precision needed to preserve GNR electronic properties \cite{cai_atomically_2010,ruffieux_electronic_2012,ruffieux_-surface_2016}. Doping of GNRs is a key aspect to fully develop the possibilities of these nanostructures as alternative material for semiconductor applications. This bottom-up strategy for growing GNRs does not only allow tuning the electronic structure of the ribbons via width control \cite{lipton-duffin_synthesis_2009,cai_atomically_2010,chen_tuning_2013,basagni_moleculesoligomersnanowiresgraphene_2015,zhang_-surface_2015,kimouche_ultra-narrow_2015,liu_toward_2015,talirz_-surface_2017}, but also opens the possibility to chemically dope them. The high precision of the on-surface built structures allows to understand atomistically the effect of dopants or functional groups in the electronic structure of the ribbon. Such effects can modify the band alignment \cite{cai_graphene_2014,bronner_aligning_2013,zhang_direct_2014}, change the band gap \cite{nguyen_bottom-up_2016}, modify the Density of States (DOS) of a ribbon by inducing new bands \cite{kawai_atomically_2015,cloke_site-specific_2015} or generate highly reflective electron scatterers \cite{carbonell-sanroma_quantum_2017}.

Up to now, the most common approach to dope GNRs has been the chemical substitution of carbon atoms by heteroatoms in the organic precursor \cite{cai_graphene_2014,bronner_aligning_2013,zhang_direct_2014,nguyen_bottom-up_2016,kawai_atomically_2015,cloke_site-specific_2015}. However, the on-surface synthesis strategy provides further tuning flexibility, such as the addition of functional groups to the GNR structure. The large variety of functional groups compatible with the synthesis of molecular precursors potentially adds a huge versatility to GNRs. For example, alkyne functional groups could be used as reaction centers for further on-surface reactions, such as Sonogashira \cite{sanchez-sanchez_sonogashira_2015} or Glaser \cite{gao_effect_2013} couplings, that could result in precise two dimensional GNR networks. Additionally, selected functional groups could be used to attach optically active centers to the GNR, such as fluorophores \cite{chong_narrow-line_2016,chong_ordinary_2016} . In terms of electronic band adjustments, electron donor or withdrawal groups can dope the electronic structure of GNR while being preferential sites for the coordination of transition metal atoms.

Here we report the synthesis on a Au(111) surface  of 7-armchair graphene nanoribbons (7-AGNRs) with nitrile (CN) groups substituted at the edges, which are of special interest due to their strong electron acceptor behavior. The on-surface reaction of cyano substituted dibromo bianthracene precursors ( \textbf{3}, Figure \ref{CNchem}) results in the formation of 7-AGNRs with CN groups protruding from the bay regions of the AGNR. We use high resolution Scanning Tunneling Microscopy (STM) imaging to obtain reliable information on the products of the reaction, such as the detachment of some CN functional groups during the reaction process and the on-surface formation of pyridine rings. Moreover, by means of Scanning Tunneling Spectroscopy (STS) and Density Functional Theory (DFT) calculations we show that CN functionalization induces a downshift on the ribbon bands of $\sim$0.3 eV per CN added. 

\section{Results and discussion}

\begin{figure}
\centering
\includegraphics[scale=0.85]{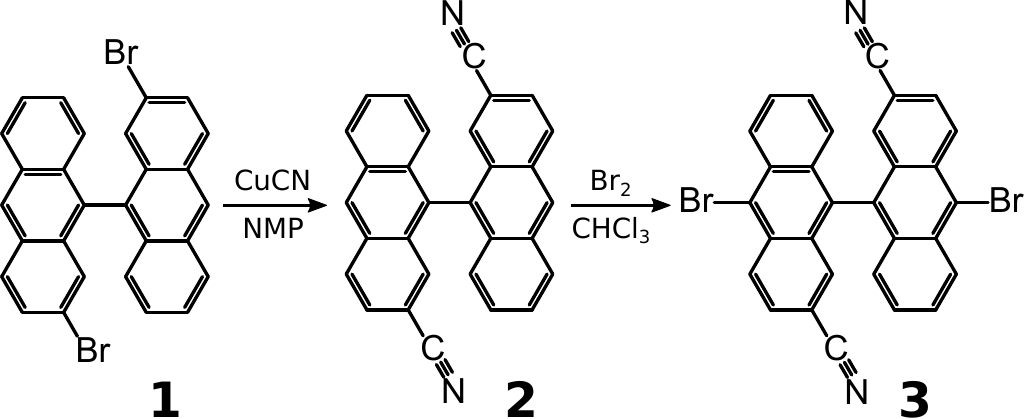}
\caption{Synthetic process to obtain cyano-substituted dibromobianthracene precursors \textbf{3} from 2,2'-dibromo-9,9'-bianthracene (\textbf{1}).} \label{CNchem}
\end{figure}

In order to introduce the CN groups in our 7-AGNR we first synthesized precursor \textbf{3} from dibromobianthracene \textbf{1} \cite{de_oteyza_substrate-independent_2016}, as depicted in Figure \ref{CNchem}. Treatment of 2,2'-dibromo-9,9'-bianthracene (\textbf{1}) with CuCN substituted the Br atoms in compound \textbf{1} for CN groups present in bianthracene \textbf{2}. Then 10,10'-dibromo-[9,9'-bianthracene]-2,2'-dicarbonitrile (\textbf{3}) was subsequently produced by means of regioselective bromination of compound \textbf{2} (see methods).

\begin{figure} [hb!]
\centering
\includegraphics[scale=0.85]{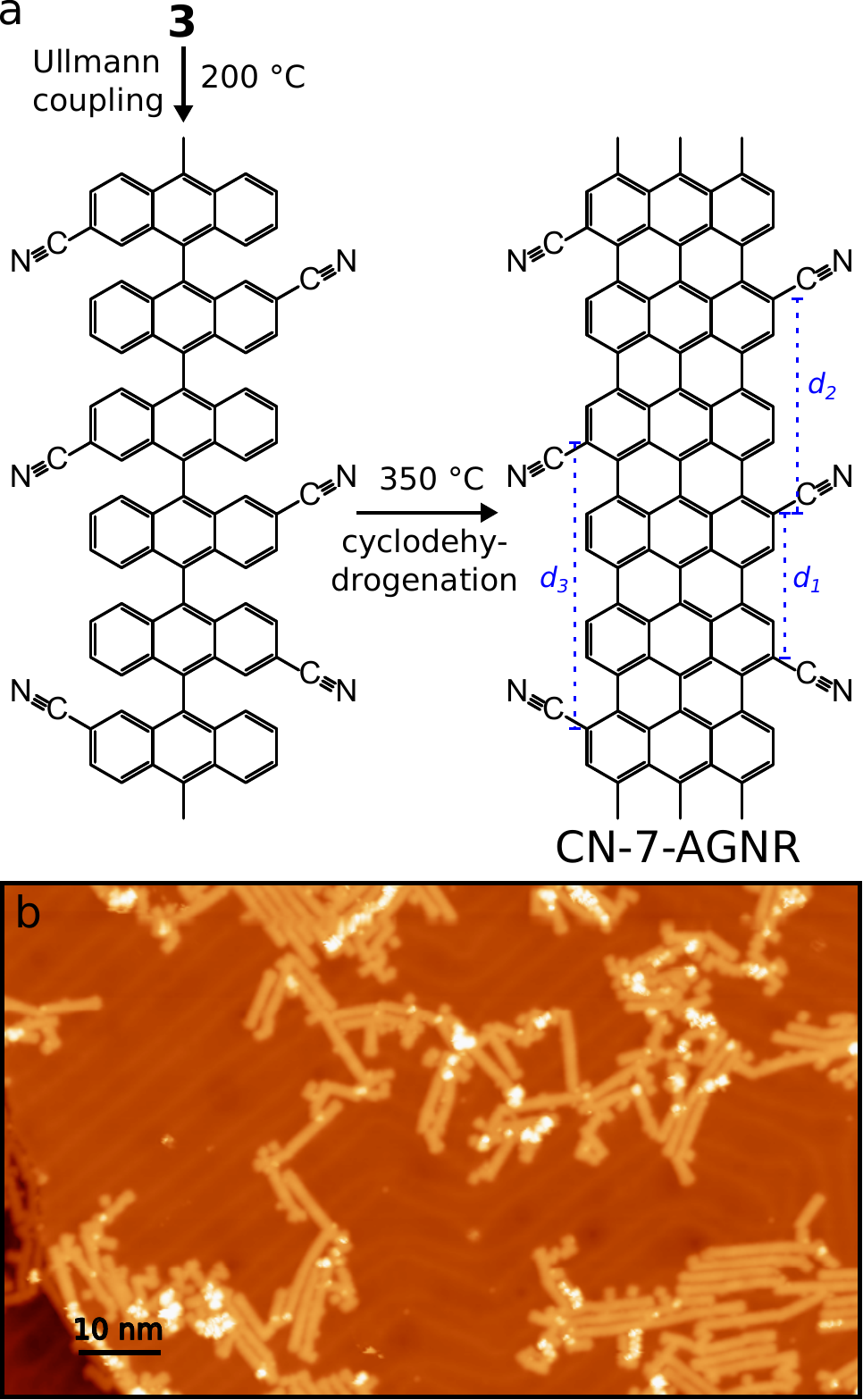}
\caption{(a) On-surface GNR preparation. The polymerization of compound \textbf{3} and the subsequent planarization yield CN-7-AGNRs, with different inter-cyano distances, indicated as: $d_1$ = 5.6 \AA, $d_2$ = 8.4 \AA\, and $d_3$ = 11.2 \AA. (b) Overview STM image of the reacted CN-7-AGNR (\Vs=1.2 V, \It=140 pA).} \label{CNgrowth}
\end{figure}

Molecular precursor \textbf{3} was sublimated at $230\,^{\circ}\mathrm{C}$ from a Knudsen cell onto a Au(111) substrate kept at room temperature. Figure \ref{CNgrowth}a depicts the reaction process leading to CN functionalized 7-AGNR. Following the deposition step, the sample was annealed to  $200\,^{\circ}\mathrm{C}$ for 10 minutes in order to induce the polymerization of monomer \textbf{3} by Ullmann coupling. Finally, the sample was annealed for 30 seconds at  $350\,^{\circ}\mathrm{C}$ in order to trigger the final cyclodehydrogenations. It is worth noting that precursor \textbf{3} can adopt two different prochiral configurations when confined to the flat surface, which are mantained during the polymerization step. As a result, the ribbons possess an intrinsic disorder in the CN groups distributions, leading to three possible inter CN distances (Figure \ref{CNgrowth}a). We have not observed any conformation where two CN share the same bay region, probably because  the high steric repulsion of this substitution pattern would evolve in the fragmentation of one of the CN groups on the bay region (see Figure S1 in Supporting Information for a possible mechanistic proposal).

\begin{figure*} [ht!]
\centering
\includegraphics[scale=0.8]{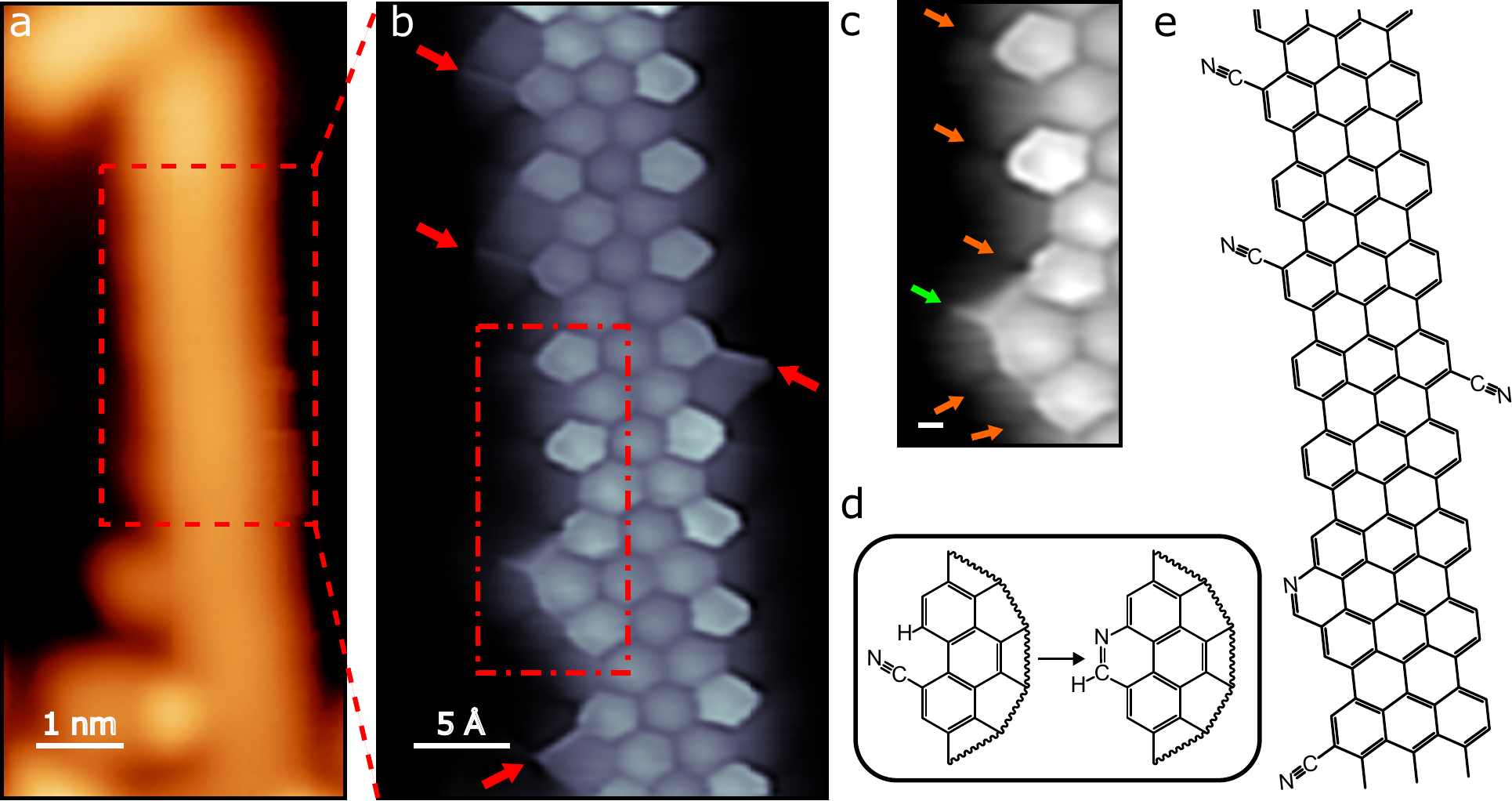}
\caption{(a) Small-scale image of CN-7-AGNR acquired with a metallic tip (\Vs=1 V, \It=50 pA). The ribbon edges look almost featureless. (b) Constant height \didv\,map with a CO-functionalized tip of the dashed region in panel \textit{a}. CN groups (red arrows) appear as linear bright protusions (\Zoff=40 pm \vs=5 mV, open feedback conditions: \Vs=5 mV \It=120 pA, V$_{\textrm{rms}}$=10 mV \textit{f}=699 Hz). (c) Zoomed in image of the dot-dashed region in panel \textit{b} highlighting the additional pyridine ring. Orange arrows point the typical dark contrast around carbon atoms saturated with hydrogen. Green arrow points the bright contrast of the  edge nitrogen atom. Scale bar corresponds to 1 \AA.(d) Reaction scheme of the nitrile cycloisomerization into pyridine rings.  (e) Chemical structure of the ribbon in panel \textit{b}.} \label{CNCO}
\end{figure*}

Figure \ref{CNgrowth}b shows an STM overview of the resulting ribbons. The typical lenghts are around 8 to 10 nm, much shorter than average lengths for pristine 7-AGNR \cite{cai_atomically_2010}. Additionally the ribbons appear aligned in arrays in contrast with the disperse arrangement of pristine 7-AGNR \cite{cai_atomically_2010}. This points towards the presence of nitrile groups in the ribbons, which drive the clustering of the GNRs via van der Waals forces, the electric dipoles created by the nitrile groups (see below) in the edges of neighboring ribbons can also contribute to this attractive inter-ribbon interaction with the appropriate relative orientation.

However, the identification of the CN-functionalized sites in the STM images is difficult. To overcome this limitation we used a CO-functionalized tip \cite{gross_chemical_2009,kichin_single_2011}. As detailed in the section Methods, this allows STM imaging of GNRs with intramolecular resolution. Figure \ref{CNCO}b shows a \didv\,image of a section of a ribbon (indicated in Figure \ref{CNCO}a) obtained using a CO-functionalized tip, resolving clearly its graphenoid backbone structure and additional features at the edges. The high resolution image  allows us to characterize with high precision the edges of the ribbon and their precise functionalization. CN groups (marked with red arrows in Figure \ref{CNCO}b) appear as linear features at the edges pointing along distinctive directions. We attribute the shadow on the CN groups to the potential landscape between CN group and the close-by H in the same bay region, to which the CO tip is sensitive \cite{hapala_mechanism_2014,hamalainen_intermolecular_2014}.

The high resolution \didv\,images also highlight unexpected chemical processes occurring during the on-surface GNR synthesis. First, we found that a large fraction of the CN groups (about 50\%) are missing from the edges, probably lost during the GNR formation steps by $\sigma$-bond cleavage. Second, we observed the occasional appearance of additional rings at the edges of the 7-AGNR backbone, with a similar contrast as other six-member carbon rings (Figure \ref{CNCO}c). Moreover, these new rings present a characteristic shape at the outermost edge atoms. We propose that these new rings are pyridine rings produced through the on-surface nitrile cycloisomerization of CN groups located on GNR bay regions (Figure \ref{CNCO}d).  This on-surface cyclization, to date unreported to the best of our knowledge, is related with the copper-catalyzed synthesis of phenanthridine derivatives from biaryl-2-carbonitriles and Grignard reagents by solution chemistry \cite{zhang_copper-catalyzed_2010} (see Figure S1 in Supporting Information for a possible mechanistic proposal). The resulting product can be corroborated by the high resolution dI/dV images, where the new H on the pyridine ring appears as a darker shadow, exactly as other aromatic hydrogen atoms nearby (orange arrows in Figure \ref{CNCO}c). In contrast, a bright line in the images points to the position of the pyridinyl nitrogen atom of the heterocycle (green arrow in Figure \ref{CNCO}c).  

\begin{figure} [t!]
\centering
\includegraphics[scale=0.85]{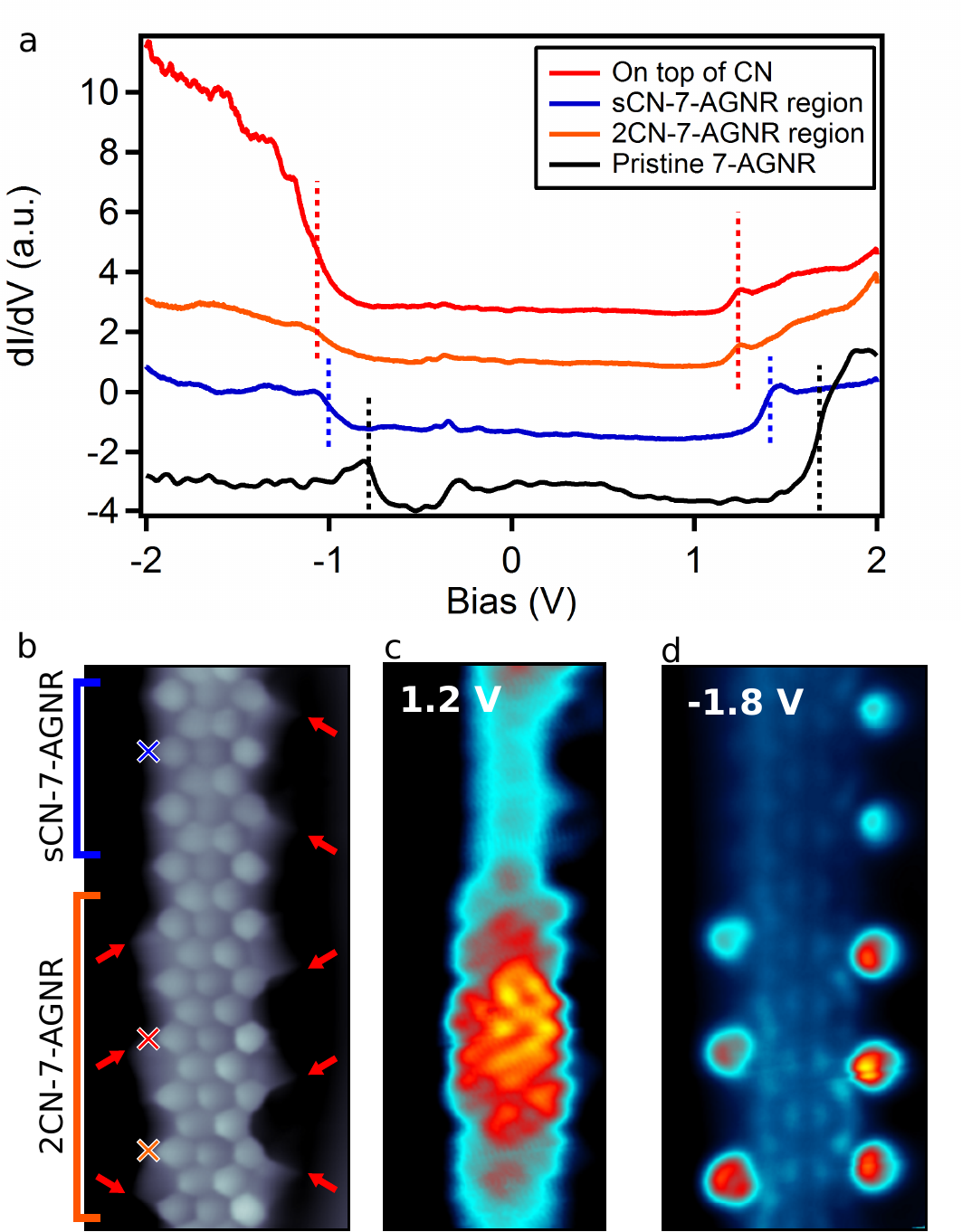}
\caption{(a) Differential conductance (\didv) spectra taken at different sites of the ribbon in panel \textit{b}. Red spectrum taken on top of a CN group. Orange spectrum taken between CN groups in the cyano rich regions (2CN-7-AGNR). Blue spectrum taken on the ribbon region with CN groups only on one edge (sCN-7-AGNR). A pristine 7-AGNR spectrum is shown in black for comparison. Dashed lines indicate the CB+1 and VB onsets for the different spectra. Onsets in sCN-7-AGNR regions are downshifted 0.28 eV respect to the pristine 7-AGNR. Onsets on the 2CN-7-AGNR region are downshifted an additional $\sim$0.18 eV for the CB+1 and $\sim$80 meV for the VB (open-feedback parameters CN-7-AGNR spectra: \Vs=1 V, \It=400 pA, V$_{\textrm{rms}}$=14 mV \textit{f}=699 Hz, open-feedback parameters pristine 7-AGNR spectrum: \Vs=0.5 V, \It=75 pA, V$_{\textrm{rms}}$=12 mV \textit{f}=700 Hz). (b) Constant height \didv\,image with a CO-functionalized tip. Red arrows mark the positions of the CN groups. Colored crosses mark the position where the spectra in panel \textit{a} were taken. sCN-7-AGNR and 2CN-7-AGNR regions are highlighted in blue and orange respectively. (\Zoff=40 pm, \vs=5 mV). (c) Constant height \didv\,map of the ribbon in panel \textit{b} taken at \Vs=1.2 V. The CB+1 onset is further downshifted in CN rich regions (\Zoff=-90 pm). (d) Constant height \didv\,map of the ribbon in panel \textit{b} taken at \Vs=-1.8 V. The map shows the large contribution of the CN groups to the states at this energy induced by the CN groups (\Zoff=-140 pm).  Open feedback conditions for CO \didv\,maps: \Vs=5 mV \It=110 pA, (b) V$_{\textrm{rms}}$=12 mV; (c,d) V$_{\textrm{rms}}$=20 mV, \textit{f}=699 Hz.)} \label{CNSTS}
\end{figure}

We performed STS measurements to investigate the impact of the CN edge functionalization on the electronic structure of the 7-AGNR (Figure \ref{CNSTS}a). Figure \ref{CNSTS}b shows a high resolution image of a GNR section, showing two regions with different density of remaining CN groups (marked with red arrows). One of the regions maintains the CNs only in one of the sides (labeled as sCN-7AGNR), while the other have intact all CNs at the expected sites (2CN-7AGNR). The dI/dV spectra in these regions show two steps at 1.4 (1.2) V and -1.0 (-1.1) V for the sCN (2CN) segments. In analogy to the case of pristine GNR,  these steps are attributed to the onset of the second conduction band (CB+1) and of the valence band (VB), respectively \cite{sode_electronic_2015,talirz_-surface_2017} (see Figure S2 in Supporting Information). Comparing these   spectra  with that of a pristine GNR segment (included in Figure \ref{CNSTS}a)  we prove that the CN functionalization produces a rigid  downshift of the frontier bands, which amounts to $\sim$0.3 eV in sCN sections and $\sim$0.4 eV in 2CN regions. This behavior indicates that CN groups behave as $n$-dopants, as found for other nitrogen doped GNRs \cite{bronner_aligning_2013,cai_graphene_2014}. The spectra also indicate that the density of CN groups affects the doping strength  and,  consequently, the bands' downshift.   This is pictured in (constant heigth) dI/dV maps measured at 1.2 V (Figure \ref{CNSTS}c), which show a significantly larger dI/dV signal in the cyano-richer regions (the 2CN segments) due to the larger downshift of the CB+1 band. 

In addition to the doping of the GNR, the CN moieties lead to a sizable  accumulation of density of states in their proximity. dI/dV maps at -1.8 V (Figure \ref{CNSTS}d), a bias value well below the VB onset, find an increased conductance signal appearing mostly over the CN groups. This might suggest the existence of an impurity state similar to those observed in previous works \cite{cervantes-sodi_edge-functionalized_2008} for amine and single nitrogen edge substitution. However, the calculations presented below seem to suggest that this signal should come from a rather flat band of the ribbon that strongly hybridizes with the CN group. Furthermore, dI/dV point spectra over the CN groups (red spectrum in Figure \ref{CNSTS}a), reproduce the larger occupied DOS at CN sites, but appear as a broad background, rather than a well-defined resonance.  

The spectra also indicate a slight reduction of the bandgap upon addition of CN groups. Comparing to pristine 7-AGNR, the bandgap in sCN-7-AGNR and 2CN-7AGNR sections is $\sim$50 meV and $\sim$100 meV smaller, respectively. The bandgap closing is consistent with the increase of effective width of the $\pi$-network, since the CN groups extend the conjugation of the 7-AGNR backbone.

\begin{figure} [t!]
\centering
\includegraphics[scale=0.85]{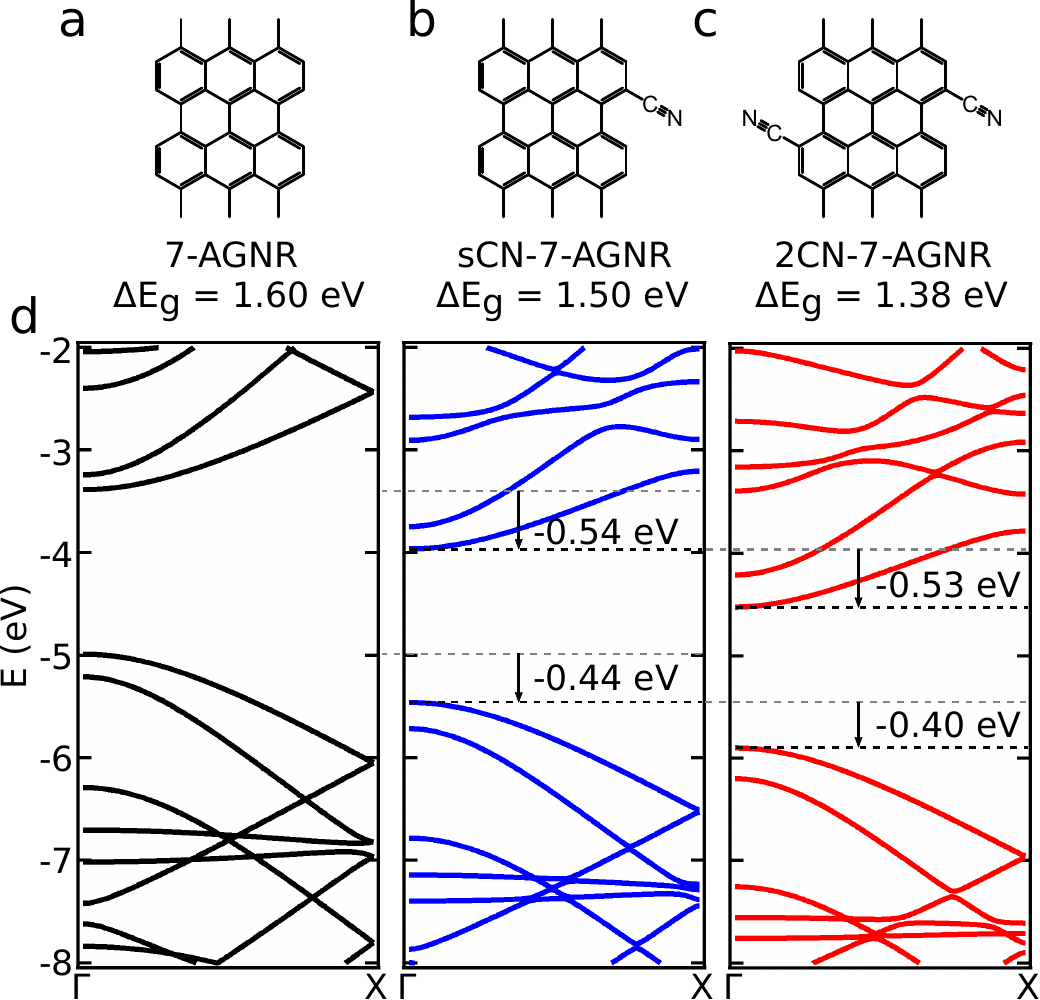}
\caption{(a-c) Periodic unit cell for the calculated freestanding ribbons: pristine 7-AGNR, sCN-7-AGNR and 2CN-7-AGNR. (d) Band structure for each different ribbon. The CB and VB show a downshift and a slight closing of the gap unpon adding CN groups. The CB+1 downshifts 0.5 eV and 0.46 eV more in sCN-7-AGNRs and 2CN-7-AGNRs, respectively. The energies of the band structures are referred to the vacuum energy.} \label{CNDFT}
\end{figure}

To complement the experimental picture on the impact of CN functionalization we performed first-principles calculations via DFT of freestanding 7-AGNRs with periodically arranged CN functional groups in one (Figure \ref{CNDFT}b, sCN-7AGNR) or both sides (Figure \ref{CNDFT}c, 2CN-7AGNR). Figure \ref{CNDFT}d compares the calculated band structure of pristine, sCN, and 2CN  nanoribbons. The calculations reproduce both the additive downshift of the bands and the bandgap reduction upon addition of CN groups. The results show that  for each  CN  added  to the pristine ribbon, the  VB downshifts $\sim$0.4 eV, while the CB and CB+1 move $\sim$0.5 eV. In our experiments, the observed band shifts are however smaller, between  0.2 eV and 0.3 eV. We attribute this to the interaction effects of the ribbon with the metal surface, such as the screening provided by the substrate. Given the calculated band downshifts, the band gap of the functionalized ribbons closes 100 meV for a sCN-7-AGNR and 220 meV for a 2CN-7-AGNR which agrees qualitatively with the experimental results. The origin of the bandgap closing is the extension of the conjugated $\pi$-network due to the addition of the CN moieties. To prove this, we calculated the band structure of a similar 7-AGNR substituting the CN groups by acetylene groups, which extend the conjugate system in a similar way (2CCH-7-AGNR, Figure S3 in Supporting Information). Our results show a similar bandgap ($\Delta E_g$=1.36 eV) to that of 2CN-7-AGNR, although without a downshift of the VB in this case, confirming our previous hypothesis.

The rigid downshift of the bands is an observed trend after the incorporation of electronegative species onto GNRs \cite{bronner_aligning_2013,cai_graphene_2014}. However, the band's downshifts induced by CN groups are larger than those induced by nitrogen heterocycles. Even though the ratio between carbon and nitrogen per ribbon cell is smaller in 2CN-7-AGNR, our DFT results find a 0.5 eV downshift per CN group compared to the 0.13 eV per edge nitrogen in chevron GNRs \cite{cai_graphene_2014}. Thus, our results indicate that cyano moieties behave as more efficient n-dopants than nitrogen heterocycles.

\begin{figure} [ht!]
\centering
\includegraphics[scale=0.5]{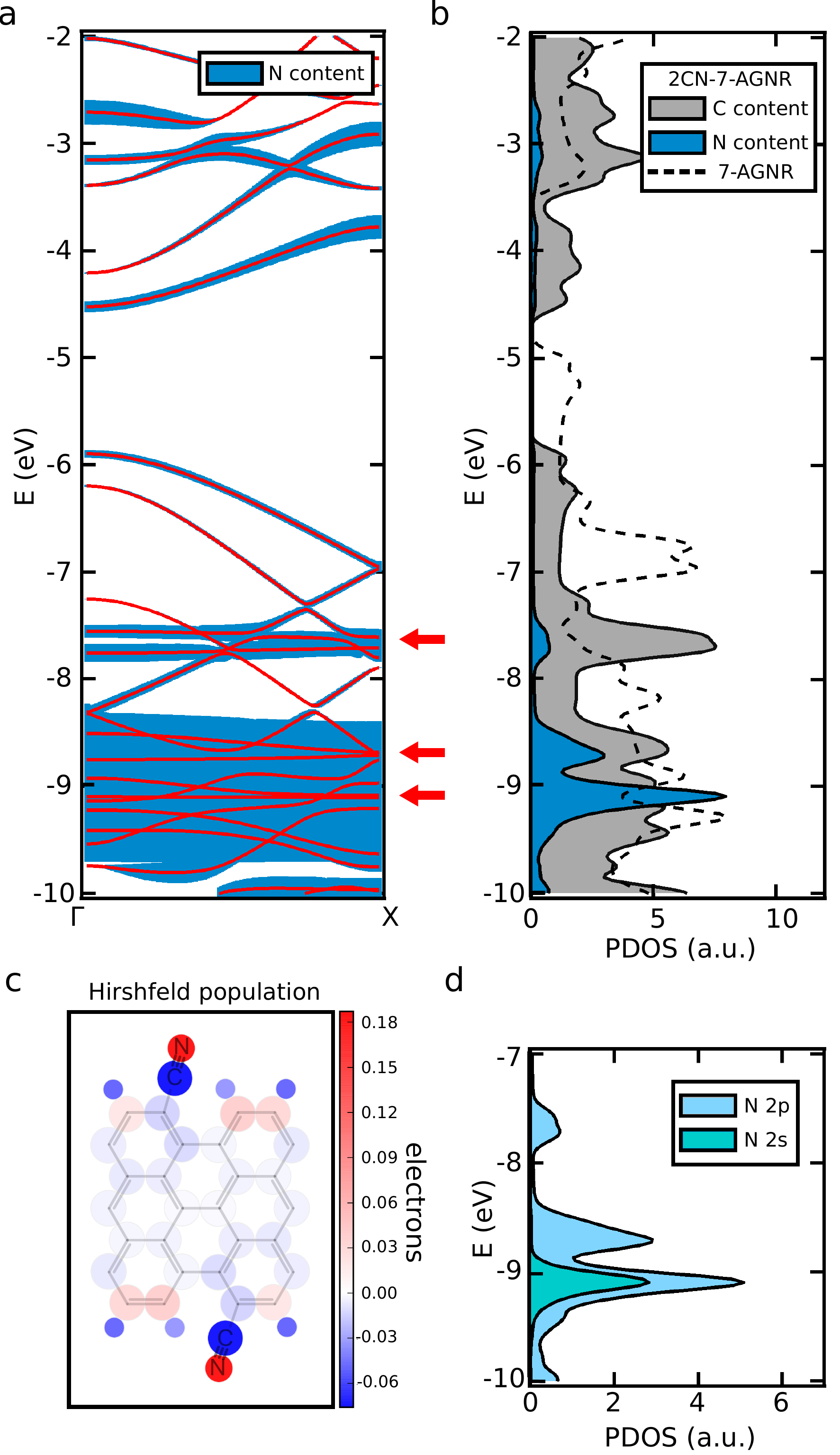}
\caption{(a) Band structure of a 2CN-7-AGNR showing the nitrogen content of the bands. The CB and the VB show a clear hybridization with the CN groups. Flat bands with strong Nitrogen weight are observed close to -7.7 eV and -9 eV (red arrows). The zero energy of the band structures is referred to the vacuum energy. (b) DOS projected on C and N atoms of a 2CN-7-AGNR (grey, blue) and a pristine 7-AGNR (dashed line). The CB and VB show smaller N contributions compared with the states associated with the flat bands at  $\sim\,$-7.7 eV and $\sim\,$-9 eV. (c) Hirshfeld population plot of a 2CN-7-AGNR (see methods). The nitrogen atoms accumulate 0.18$e$, while the carbon network is slightly depleted of charge. (d) Orbital contributions to the nitrogen PDOS of the strongest nitrogen peaks in panel \textit{b}. The peak at $\sim\,$-7.7 eV shows only $p$ component, whereas the peak at $\sim\,$-9.2 eV has a strong $s$ component too.}\label{CNfatbands}
\end{figure}

To unravel the mechanism behind the band downshift induced by the CN groups we next focus on the details of interaction between these functional groups and the ribbon. Figure \ref{CNfatbands}a shows the 2CN-7-AGNR bands, with the amount of N-contribution represented by the thickness of a blue shadow. The plot shows that the the N character is widespread in the whole band structure. The origin of such strong mixing is the resonant character of the conjugation between CN and 7-AGNR mesh. The DOS is enhanced particularly at $\sim$-7.7 eV due to nearly flat bands (top red arrow, Figure \ref{CNfatbands}b) with strong $p$ character (Figure \ref{CNfatbands}d), presumably being responsible to the dI/dV enhancement found in the spectra of Figure \ref{CNSTS}a and in the dI/dV map of Figure \ref{CNSTS}d. Moreover, states with strong N character are found at lower energy, deep inside the filled states of the ribbon ($\sim\,$-9 eV, Figure \ref{CNfatbands}b). These states, not reached in our STS spectra, have both \textit{s} and \textit{p} character (Figure \ref{CNfatbands}d), in agreement with the nitrogen $sp$ orbital hosting the lone pair.  Since the localized lone pairs do not participate in the conjugation of the GNR $\pi$-system, these states are regarded as the impurity levels induced by the CN groups.

Most importantly, the strong electron withdrawing character of CN induces a charge redistribution over the whole ribbon (Figure \ref{CNfatbands}c) and results in sizeable dipoles at the CN sites (estimated as $\sim$-2.9D, see Figure S4 in Supporting Information) pointing towards the ribbon backbone. The localized dipoles generate an electrostatic potential background approximately 1 eV lower than in pristine 7-AGNRs, which turns the ribbon more electronegative, and is consistent with the observed band downshifts (see Figure S5 in Supporting Information). Thus, our results suggest that the bands downshift is a consequence of the charge redistribution induced by the CN groups.

\section{Conclusions}

In summary, we have shown the growth of cyano substituted 7-AGNR and studied the impact of nitrile functional groups on the electronic structure of the ribbons. The CN groups increase the reactivity of the 7-AGNRs as inferred from the bunching of the ribbons into aligned clusters and the overall shorter ribbon lengths. STM imaging with CO-functionalized tips resolves with high precision the intramolecular structure of the ribbons and finds a significant loss of CN groups, close to 50\%. Moreover, the high resolution images unveil the on-surface nitrile cycloisomerization into pyridine rings, which has not been previously reported.

By combining STS and DFT calculations we have shown that the addition of CN groups conjugated to the GNR structure reduces its bandgap, since the $\pi$-network of the GNR is extended. Furthermore, we have demonstrated that CN groups behave as $n$-dopants. The CN groups downshift the bands $\sim$0.5 eV per CN added ($\sim$0.3 eV in the experiments), which is significantly more than the doping observed in substitutional nitrogen edge atoms. The downshift of the bands stems from the strong electron withdrawing character of nitrile groups, which induce dipoles at the CN sites. The charge redistribution causes a downshift of the electrostatic potential in the GNR backbone, resulting in an increased ribbon electronegativity. In conclusion, our work shows the potential of using functional groups as tools to modify the physicochemical properties of GNRs, albeit further work concerning the stability of said groups is needed.

\section{\label{sec:level1}Methods}

Our experiments were performed in  custom designed low temperature STM, under ultra-high vacuum, and
at 5 K. Prior to the monomer deposition, the Au(111) single crystal surface was sputtered with Ne$^{+}$ ions, typically for 10 min and then annealed in UHV at temperatures between 490 and 500$\,^{\circ}\mathrm{C}$, around 10 min. Gold coated tungsten tips were used for imaging and spectroscopy. CO functionalized tips were obtained by picking up a CO molecule adsorbed on top of NaCl islands deposited after the on-surface reaction was done. The NaCl was deposited to simplify the CO recognition and pickup. We measured the differential conductance spectra and maps by applying a small modulation to the sample bias and using the lock-in technique to obtain a signal proportional to \didv\, from the first harmonic of the tunneling current.

High resolution images with CO-functionalized tips were obtained in constant height mode, at biases close to Fermi energy and recording the differential conductance signal. Open feedback conditions are given on top of the ribbon. Imaging close to the Pauli repulsive regime allows to obtain intramolecular resolved images \cite{temirov_novel_2008,weiss_imaging_2010,kichin_single_2011,kichin_calibrating_2013,hapala_mechanism_2014,krejci_principles_2017}
STM images were processed and analyzed using the WSxM software \cite{horcas_<span_2007}.

Details on the synthetic procedure to obtain monomer \textbf{3} and its spectroscopic characterization are given in the Supporting Information.

Our first-principles simulations where performed via DFT as implemented in SIESTA code.\cite{Soler2002} We used van der Waals density functional of Dion \emph{et al.}\cite{Dion2004} with the modified exchange by Klime\v{s}, Bowler and Michaelides\cite{Klimes2010}. A double-zeta basis set was used to expand the valence-electron wave functions, while the core electrons were described by non-conserving Troullier-Martins pseudopotentials\cite{Troullier1991}. An energy cutoff of 350 Ry was used for real space integrations and the orbital radii were defined using a 30 meV energy shift.\cite{Soler2002} We used a sampling of 50 k-points along the periodic direction. All structures were fully optimized with the conjugate gradient method until all forces were lower than 10 meV/\AA. A unit cell length of 8.64 \AA\,, 8.65 \AA\, and 8.67 \AA\, were obtained for pristine 7-AGNR, sCN-7-AGNR and 2CN-7-AGNR, respectively, in the periodic direction (Figure 5a), while in the other directions 40 \AA\, (perpendicular to the ribbon plane) and 50 \AA\ (in the ribbon plane) cell sizes were considered. To check the convergence of the calculated positions of the
electronic levels with respect to vacuum, we performed calculations
using even larger inter-ribbon distances (90 \AA\ and 100 \AA,
respectively, along the perpendicular or parallel to the ribbon plane)
and we found the highest differences to be lower than 50 meV. The atomic population analysis were performed using the Hirshfeld scheme for partitioning the electron density \cite{Hirshfeld1977}.

\section{Acknowledgements}

The authors thank D.G. de Oteyza for his insights on GNR doping. This work was supported by FP7 FET-ICT ``Planar Atomic and Molecular Scale devices'' (PAMS) project (funded by the European Commission under contract No. 610446), by the Spanish Ministerio de Econom\'ia y Competitividad (MINECO) (cooperative grant No. MAT2016-78293 and grant FIS 2015-62538-ERC) and the Basque Government (Dep. de Educaci\'on and UPV/EHU,  Grant No. IT-756-13, and Dep. Industry, Grant PI\_2015\_1\_42).

\section{Supporting Information}

Possible mechanism for CN cleavage and pyridine formation. Wavefunctions of 2CN-7-AGNR bands. Band structure of acetylene functionalized 7-AGNR and their derived bandgaps. Charge redistribution plots and edge dipole estimation. Calculation and plot of the averaged electrostatic potential. Molecular precursor synthesis and characterization.

\bibliography{Bibliography}

\end{document}


\newpage
\tableofcontents
\newpage

\addcontentsline{toc}{section}{Possible CN cleavage mechanism}
\section{Possible CN cleavage mechanism}

One of the possible mechanisms for the cleavage of CN groups (and pyridine formation) could involve the presence of two nitriles in the same bay region. The high steric repulsion of this configuration would lead to the fragmentation of one CN group. Following the cleavage, the remaining CN could undergo a cycloisomerization process.

\begin{figure}
 \centering
  \includegraphics[width=1\columnwidth]{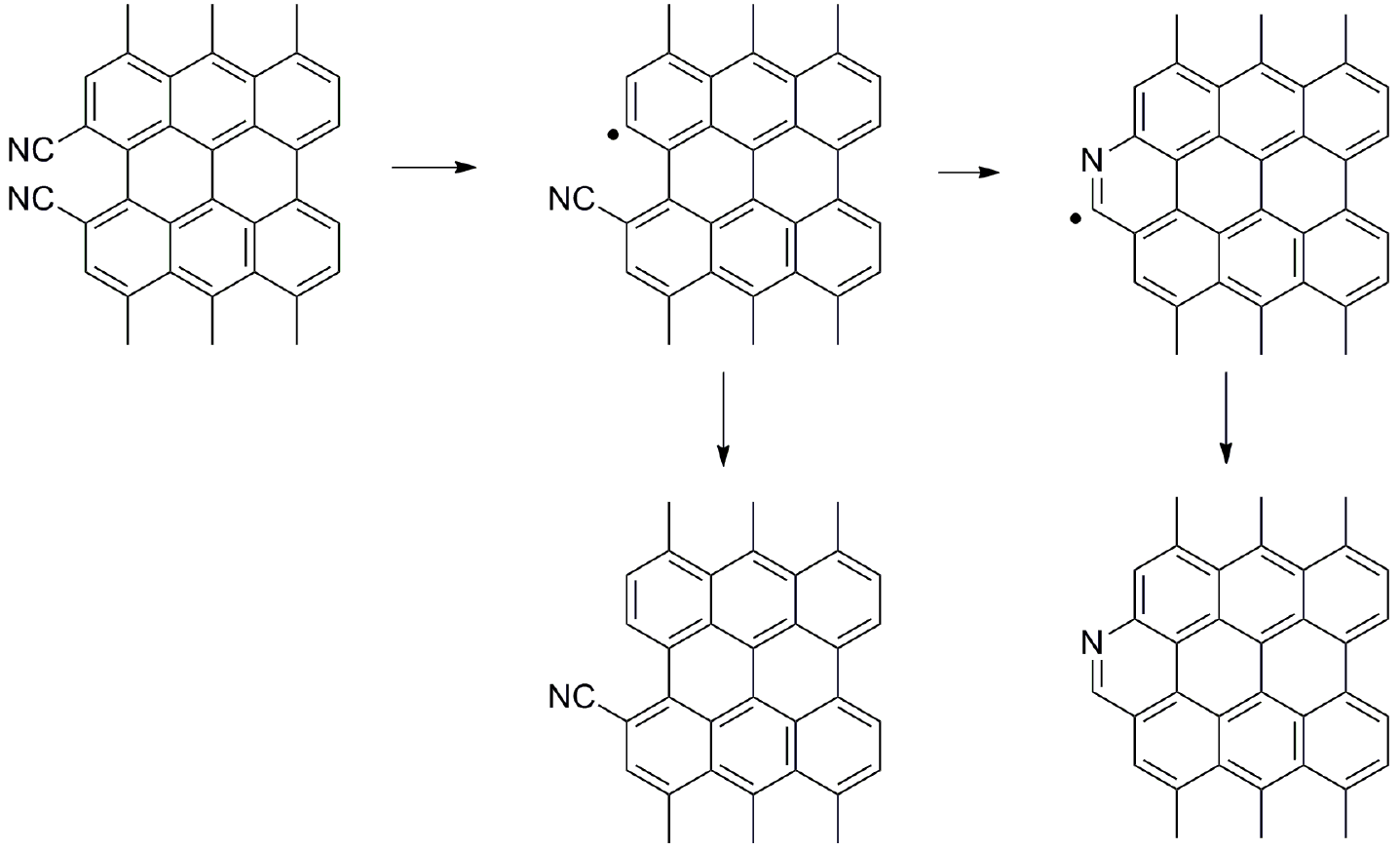}
  \caption{A possible mechanistic proposals for CN cleavage and pyridine formation.}
  \label{mechanism}
\end{figure}
\newpage

\addcontentsline{toc}{section}{DFT calculations}
\section{DFT calculations}
\addcontentsline{toc}{subsection}{2CN-7-AGNR Wavefunctions}
\subsection{2CN-7-AGNR Wavefunctions}

\begin{figure}
 \centering
  \includegraphics[scale=1]{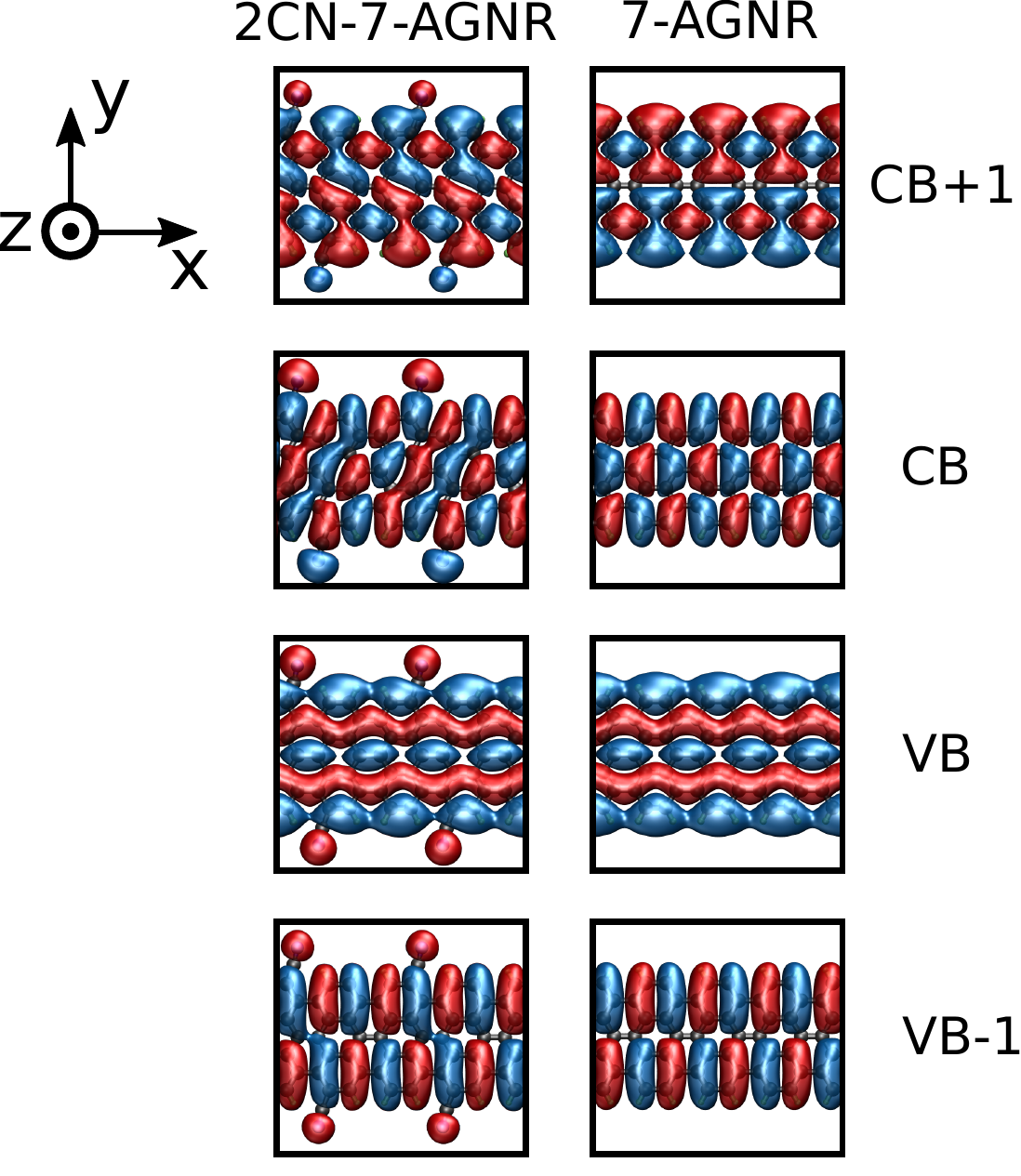}
  \caption{Real part of the wavefunctions of pristine 7-AGNR and 2CN-7-AGNR at the $\Gamma$ point. CB and VB-1 bands decay faster than CB+1 and VB along the z direction.}
  \label{wavefunctions}
\end{figure}

Figure \ref{wavefunctions} shows the real part of the bands wavefunction at the $\Gamma$ for a pristine 7-AGNR and a 2CN-7-AGNR. We note that the bands symmetry is not changed by the presence of CN groups, thus the relative intensities of the STM/STS images produced by different bands should not be altered as compared to the pristine case. As a result, the CB and VB-1 bands produce a faint signal when probed at typical tip-ribbon distances, and the STM/STS images are dominated by the CB+1 and VB bands.\cite{sode_electronic_2015,talirz_-surface_2017}

\addcontentsline{toc}{subsection}{Extension of the $\pi$-network}
\subsection{Extension of the $\pi$-network}

\begin{figure}
 \centering
  \includegraphics[width=0.5\columnwidth]{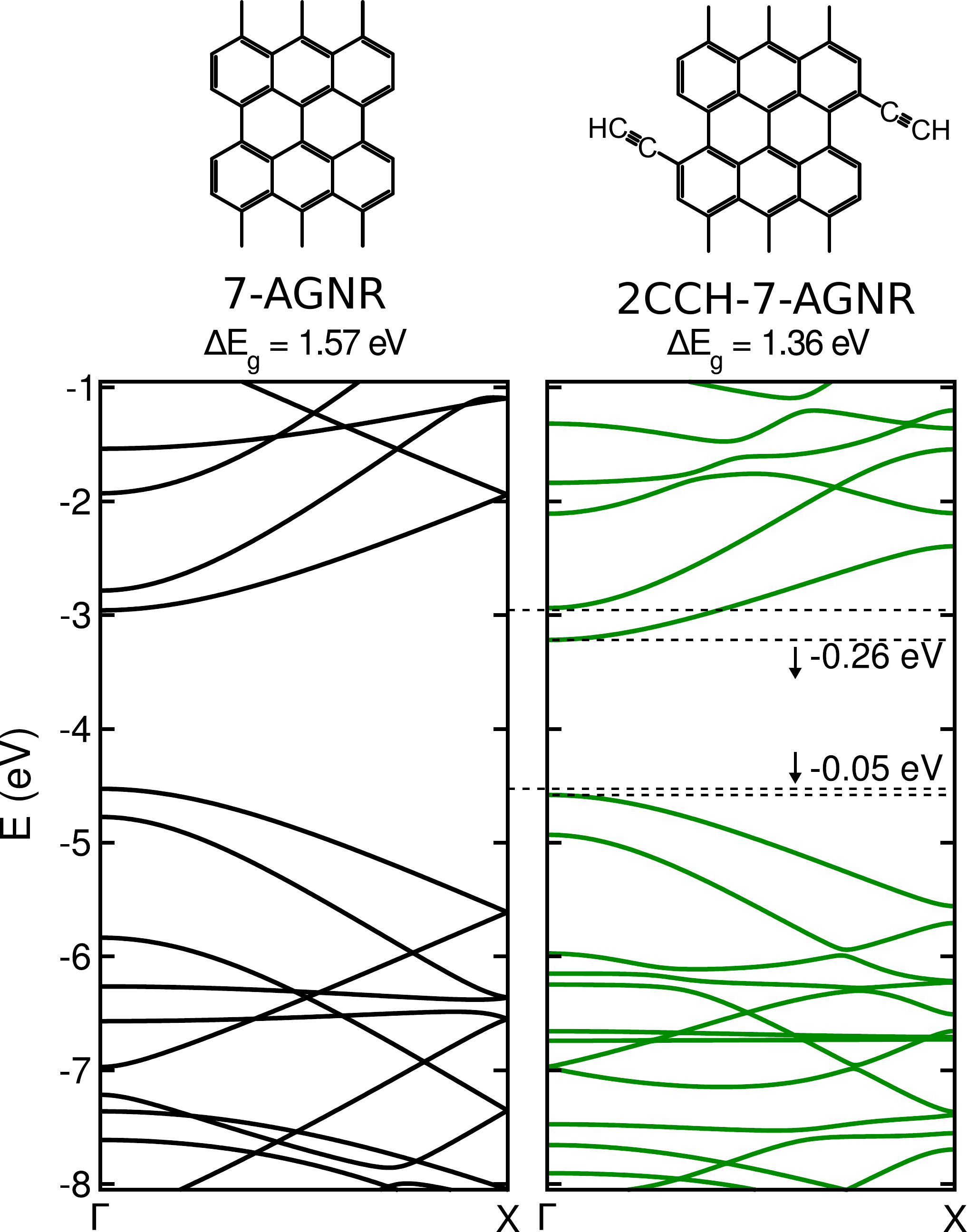}
  \caption{Band structure of a pristine 7-AGNR (left) and an acetylene functionalized 7-AGNR, 2CCH-7-AGNR (right).}
  \label{CCH7AGNR}
\end{figure}

To corroborate the origin of the band gap closing we calculated the band structure of a pristine 7-AGNR functionalized with ethylene groups in the positions of the CN groups in a 2CN-7-AGNR (Figure \ref{CCH7AGNR}). This is done to discard any major impacts of the nitrogen heteroatoms, while still maintaining the $\pi$-network extension effect. We observe a similar closing of the band gap compared to 2CN-7-AGNR ($\Delta E_g^{2CCH}$=1.36 eV vs $\Delta E_g^{2CN}$=1.38 eV). The closing occurs by the downshift of the CB, while the VB stays at almost the same energy than in the pristine case. Notice that the alignment of these band structures with respect to the vacuum level is slightly different as compared to Figure 5 in the main text. This is due to the use of a basis set with slightly more confined pseudo-atomic orbitals in this case (here we used 136 meV for the energy shift parameter\cite{Soler2002}).

\newpage

\addcontentsline{toc}{subsection}{Charge redistribution in 2CN-7-AGNR}
\subsection{Charge redistribution in 2CN-7-AGNR}

\begin{figure}
 \centering
  \includegraphics[width=1\columnwidth]{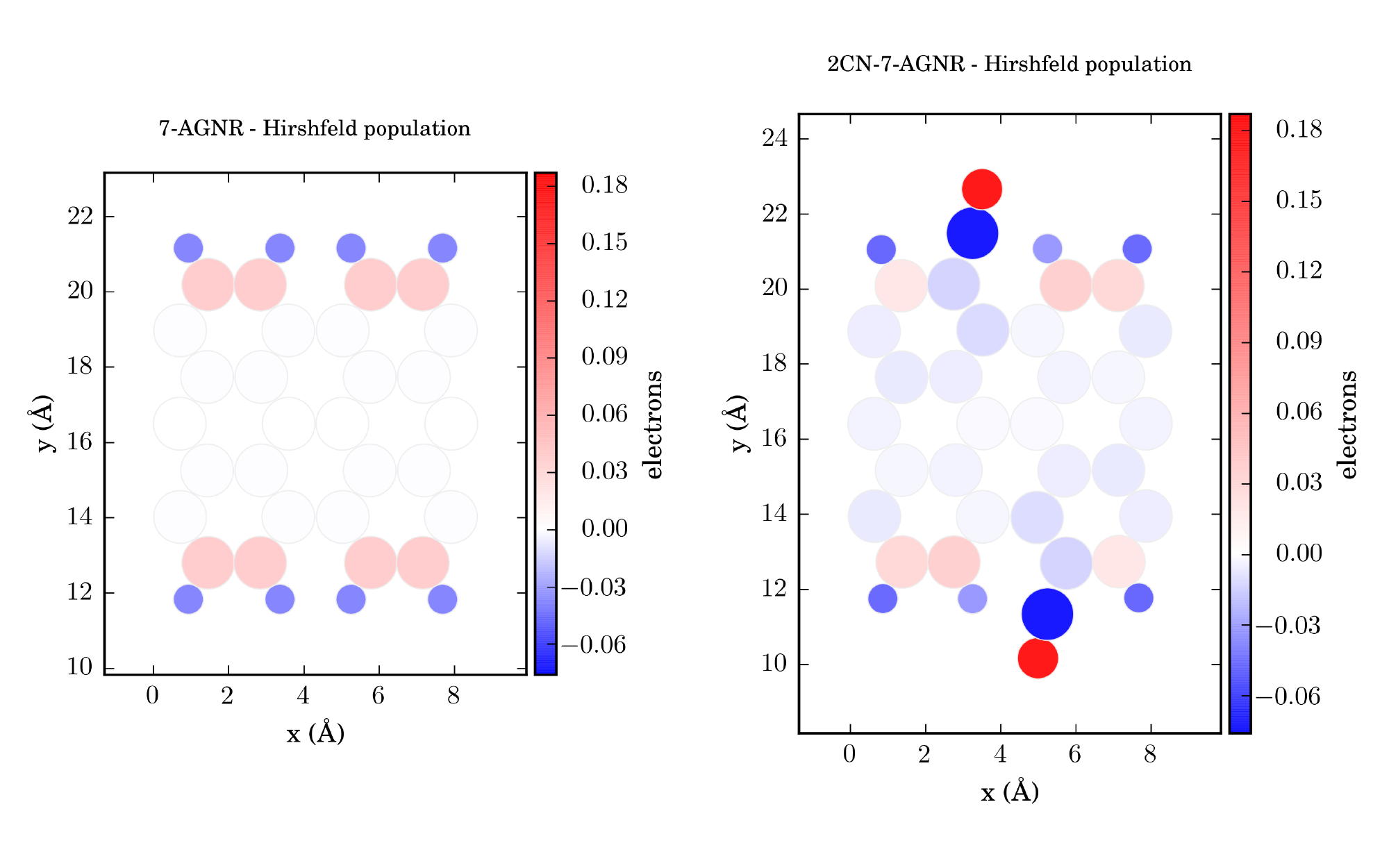}
  \caption{Hirshfeld plots showing the charge distribution of freestanding pristine 7-AGNR (left) and 2CN-7-AGNR (right).}
  \label{charge}
\end{figure}

The addition of CN groups rearranges the charge over the ribbon, as seen in the Hirshfeld plots in Figure \ref{charge} comparing the pristine and CN functionalized ribbon. To analyze the effect of the electronic density redistribution we calculated the dipole moment produced at the ribbons' edges by integrating the total valence charge density minus the sum of atomic valence charge densities (i.e. the valence charge density of each individual free atom) of half of the ribbon (from the ribbon backbone to vacuum), times the distance from the ribbon backbone:

\begin{equation}
  \boldsymbol{\mu} = \int\limits_{\text{half cell}} \left(\rho_{tot}{({\bf r})} - \sum_{I = 1}^{N_{atoms}} \rho_{I}{({\bf r})}\right) {\bf r} d{\bf r}
\end{equation}

In the pristine ribbon, each edge accumulates a dipole of 2.4D pointing from the ribbon backbone to the edge, mainly as a consequence of the charge redistribution between C-H end groups (that is 0.6D per C-H). In the case of the 2CN-7-AGNR, each edge accumulates a dipole of -1.1D (pointing to the ribbon backbone). The three C-H end groups hold similar charge population than in the pristine case. However, the CN group generates a dipole in the opposite direction. Thus we estimate that the dipole induced only by the CN is of $\mu = -1.1-0.6*3 \sim -2.9$D. This estimation is in agreement with the dipole moment of $\mu = - 2.86$D calculated for a HCN molecule.

\newpage
\addcontentsline{toc}{subsection}{Averaged electrostatic potential inside 2CN-7-AGNR}
\subsection{Averaged electrostatic potential inside the ribbon}

\begin{figure}
 \centering
  \includegraphics[width=0.75\columnwidth]{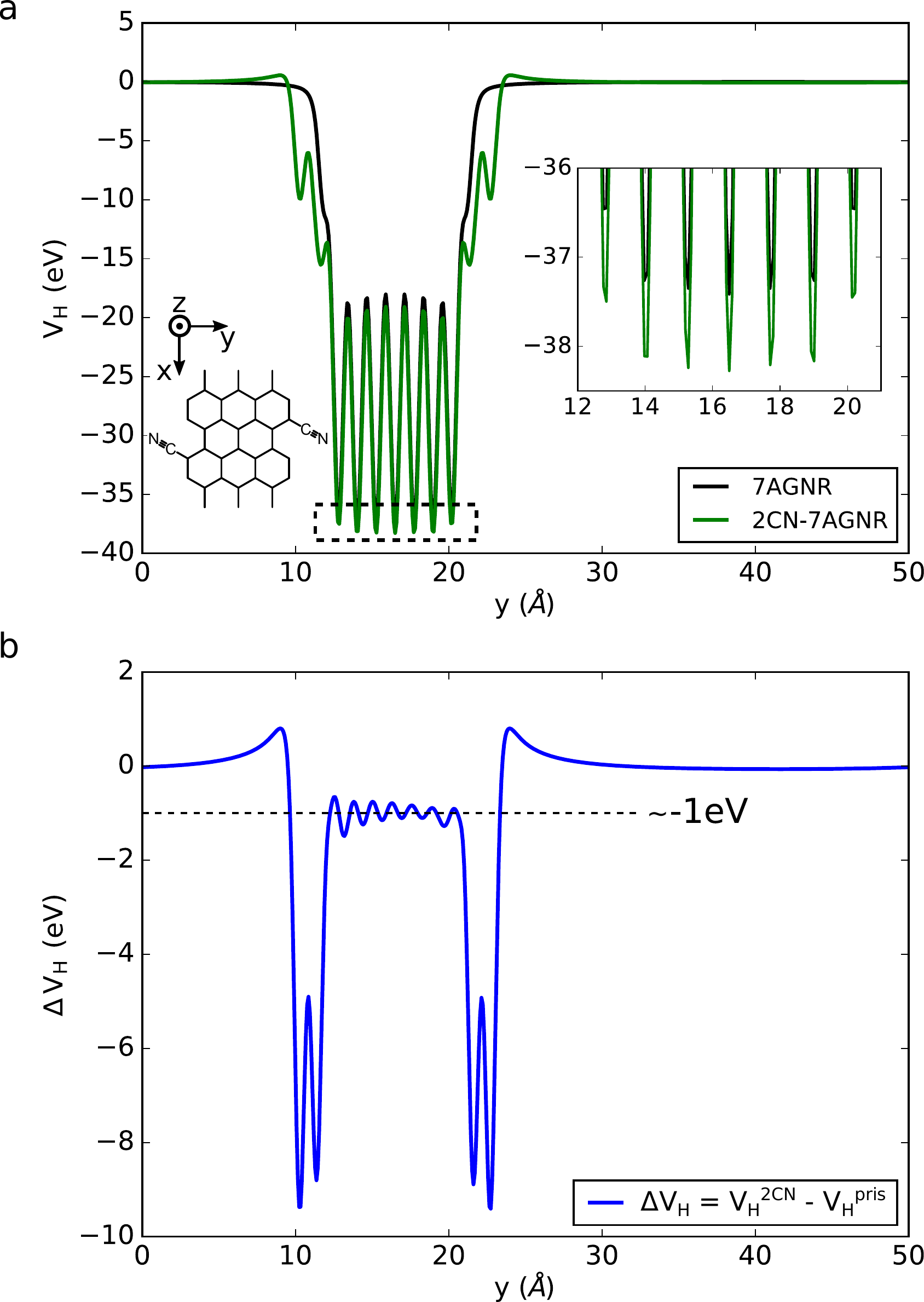}
  \caption{(a) Electrostatic potential ($V_H$) of a pristine freestanding 7-AGNR (black) and a freestanding 2CN-7-AGNR plotted at the plane of the ribbon and averaged over the periodic direction (x). Small positive shifts can be observed in the region outside the ribbon, but close to the edges, as a consequence of the CN dipoles. Inset zooms at the central region of the ribbon (dashed rectangle), where an offset of -1 eV can be seen between both cases. (b) Difference in electrostatic potential ($\Delta V_H = V_H^{2CN}-V_H^{prist}$) between both ribbons, highlighting the presence of a 1 eV downshift at the central part of the ribbon. The oscillations observed are mostly caused by small differences in atomic positions between both ribbons.}
  \label{avgpot}
\end{figure}

In order to determine the consequences of the charge redistribution caused by the CN groups, we calculate electrostatic potential of the ribbon as the sum of the Hartree potential plus the local pseudopotential.\cite{Soler2002} Figure \ref{avgpot}a shows the electrostatic potential at the plane of the ribbon (z=0) and averaged over the periodic direction (x). We clearly observe the effect of the dipoles caused by the CN groups as two $V_H>0$ shifts at the edges of the ribbon. As seen in the inset of Figure \ref{avgpot}a, the electrostatic potential at the center of 2CN-7-AGNR backbone presents a small downshift ($\sim$ -1 eV) compared to the pristine case. Figure \ref{avgpot}b shows the difference in electrostatic potential between 2CN-7-AGNR and pristine 7-AGNR ($\Delta V_H = V_H^{2CN}-V_H^{prist}$), highlighting the -1 eV electrostatic background at the backbone of the functionalized ribbon. We note as well that the shift in electrostatic background agrees nicely with the calculated band downshifts (-1.07 eV for the CB, -0.84 eV for the VB).

\newpage

\addcontentsline{toc}{section}{Molecular precursor synthesis and characterization}
\section{Molecular precursor synthesis and characterization}
\addcontentsline{toc}{subsection}{General methods}
\subsection{General methods}

\begin{figure}
 \centering
  \includegraphics[scale=1]{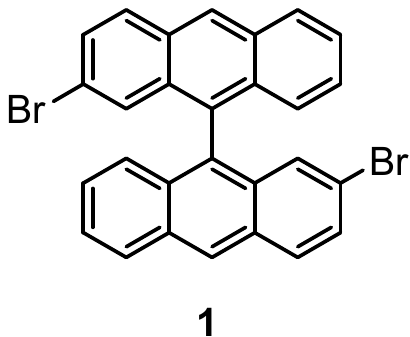}
  \caption{Chemical structure of 2,2'-Dibromobianthryl}
  \label{chiral}
\end{figure}

All reactions were carried out under argon using oven-dried glassware. TLC was performed on Merck silica gel 60 F254; chromatograms were visualized with UV light (254 and 360 nm). Flash column chromatography was performed on Merck silica gel 60 (ASTM 230-400 mesh). 1H and 13C NMR spectra were recorded at 300 and 75 MHz (Varian Mercury 300 instrument). Low-resolution electron impact mass spectra were determined at 70 eV on a HP-5988A instrument. High-resolution mass spectra (HRMS) were obtained on a Micromass Autospec spectrometer.
2,2'-Dibromobianthryl (\textbf{1}, Figure \ref{chiral}) was prepared following a published procedure \cite{de_oteyza_substrate-independent_2016}.  Commercial reagents and anhydrous solvents were purchased from ABCR GmbH, Aldrich Chemical Co., or Strem Chemicals Inc., and were used without further purification.

\addcontentsline{toc}{subsection}{Experimental details and spectroscopic data}
\subsection{Experimental details and spectroscopic data}
\addcontentsline{toc}{subsubsection}{Synthesis of [9,9'-bianthracene]-2,2'-dicarbonitrile}
\subsubsection{Synthesis of [9,9'-bianthracene]-2,2'-dicarbonitrile (\textbf{2})}

\begin{figure}[t]
 \centering
  \includegraphics[scale=1]{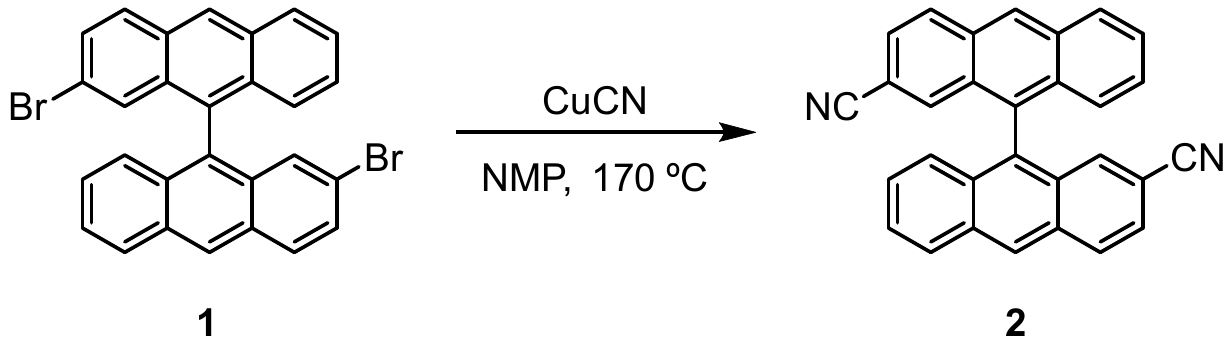}
  \caption{Processing and chemical structure of [9,9'-bianthracene]-2,2'-dicarbonitrile (\textbf{2})}
  \label{bromineless}
\end{figure}

To a flame-dried flask 2,2'-dibromobianthryl (\textbf{1}, 74 mg, 0.15 mmol), CuCN (37 mg, 0.42 mmol) and NMP (2.0 mL) were added. The resulting mixture was heated at $170\,^{\circ}\mathrm{C}$ for 20 h under inert atmosphere. After cooling, H\textsubscript{2}O (20 mL) was added and the mixture was extracted with CH\textsubscript{2}Cl\textsubscript{2} (3$\times$10 mL). The combined organic layers were dried over anhydrous Na\textsubscript{2}SO\textsubscript{4}, filtered and concentrated under reduced pressure. The crude product was purified by column chromatography (SiO\textsubscript{2}; hexane:CH\textsubscript{2}Cl\textsubscript{2} 1:1 to 1:9) to afford [9,9'-bianthracene]-2,2'-dicarbonitrile (Figure \ref{bromineless}, \textbf{2}) (38 mg, 64\%) as a green solid (m.p. $344\,^{\circ}\mathrm{C}$). \textsuperscript{1}H-NMR (300 MHz, CDCl\textsubscript{3}) $\delta$: 8.78 (s, 2H), 8.27 (d, \textit{J} = 8.8 Hz, 2H), 8.23 (d, \textit{J} = 8.6 Hz, 2H), 7.60 (d, \textit{J} = 7.6 Hz, 2H), 7.55 (m, 2H), 7.40 (s, 2H), 7.27 (m, 2H), 7.04 (d, \textit{J} = 8.9 Hz, 2H) ppm. \textsuperscript{13}C-NMR (75 MHz, CDCl\textsubscript{3}) $\delta$: 133.8 (2CH), 133.2 (2C), 133.1 (2C), 132.3 (2C), 131.3 (2C), 130.3 (2CH), 129.8 (2C), 128.8 (2CH), 128.5 (2CH), 127.6 (2CH), 127.2 (2CH), 126.5 (2CH), 124.5 (2CH), 119.1 (2C), 109.7 (2C) ppm. EM (EI) $m/z$ (\%): 404 (100), 375 (20), 187 (21). HRMS (EI) for C\textsubscript{30}H\textsubscript{16}N\textsubscript{2}; calculated: 404.1313, found: 404.1311.

\addcontentsline{toc}{subsubsection}{Synthesis of 10,10'-dibromo-[9,9'-bianthracene]-2,2'-dicarbonitrile}
\subsubsection{Synthesis of 10,10'-dibromo-[9,9'-bianthracene]-2,2'-dicarbonitrile (\textbf{3})}

\begin{figure}
 \centering
  \includegraphics[scale=1]{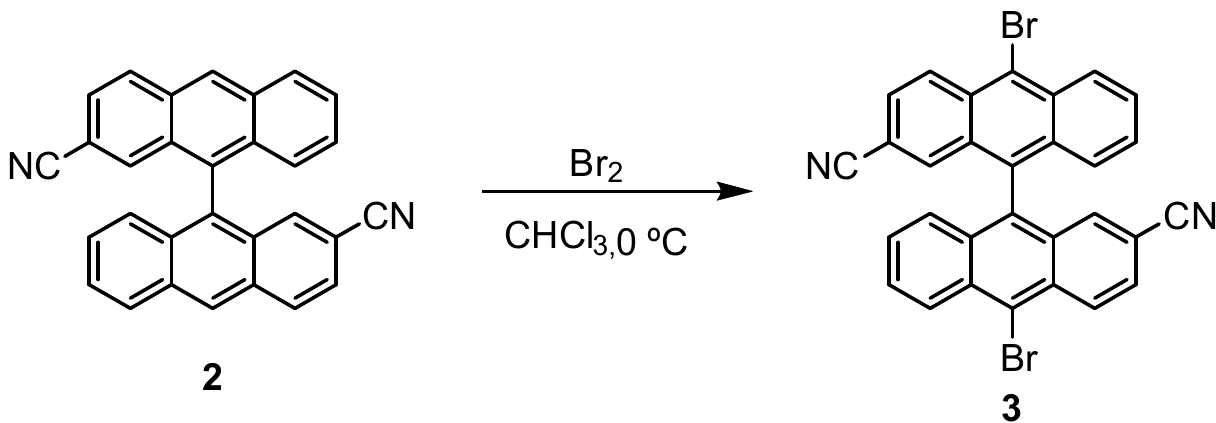}
  \caption{Processing and chemical structure of 10,10'-dibromo-[9,9'-bianthracene]-2,2'-dicarbonitrile(\textbf{3})}
  \label{cyano}
\end{figure}

To a solution of compound \textbf{2} (74 mg, 0.18 mmol) in CHCl\textsubscript{3} (5 mL) a Br\textsubscript{2} solution (4.40 mL, 0.44 mmol, 0.1 M in CHCl\textsubscript{3}) was added dropwise at 0 ºC. Then, the mixture was allowed to reach room temperature and stirred for 16h. The resulting mixture was washed with saturated solution of Na\textsubscript{2}S\textsubscript{2}O\textsubscript{3} (5 mL) and the aqueous layer was extracted with CHCl\textsubscript{3}(2x5 mL). The combined organic layers were dried over anhydrous Na\textsubscript{2}SO\textsubscript{4}, filtered and evaporated under reduced pressure. The crude product was purified by column chromatography (SiO\textsubscript{2}, hexane/CH\textsubscript{2}Cl\textsubscript{2} 1:2 to 1:4) to afford 10,10'-dibromo-[9,9'-bianthracene]-2,2'-dicarbonitrile (Figure \ref{cyano}, \textbf{3}) (80 mg, 77\%) as a yellow solid (m.p. $380\,^{\circ}\mathrm{C}$). \textsuperscript{1}H-NMR (300 MHz, CDCl\textsubscript{3}) $\delta$: 8.85 (d, \textit{J} = 9.2 Hz, 2H), 8.78 (d, \textit{J} = 9.0 Hz, 2H), 7.73 (dd, \textit{J} = 8.3, 7.3 Hz, 2H), 7.67 (dd, \textit{J} = 9.1, 0.5 Hz, 2H), 7.38 (s, 2H), 7.31 (m, 2H), 7.04 (d, \textit{J} = 8.7 Hz, 2H) ppm. \textsuperscript{13}C-NMR (75 MHz, CDCl\textsubscript{3}) $\delta$: 133.8 (2CH), 133.3 (2C), 133.2 (2C), 132.5 (2C), 130.8 (2C), 130.7 (2C), 130.5 (2CH), 129.3 (2CH), 128.9 (2CH), 128.3 (2CH), 127.1 (2CH), 126.6 (2CH), 125.7 (2C), 118.5 (2C), 110.9 (2C) ppm. EM (EI) $m/z$ (\%): 562 (100), 482 (6), 400 (31), 375 (13), 201 (17), 188 (20). HRMS (EI) for C\textsubscript{30}H\textsubscript{14}Br\textsubscript{2}N\textsubscript{2}; calculated: 559.9524, found: 559.9529.
\newpage

\addcontentsline{toc}{subsection}{\textsuperscript{1}H and \textsuperscript{13}C NMR spectra}
\subsection{\textsuperscript{1}H and \textsuperscript{13}C NMR spectra}

\begin{figure}
 \centering
  \includegraphics[width=0.9\columnwidth]{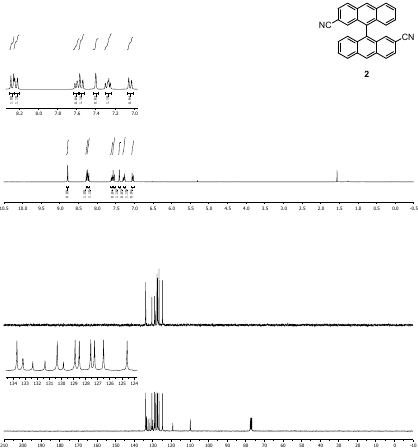}
  \caption{\textsuperscript{1}H and \textsuperscript{13}C NMR spectra of [9,9'-bianthracene]-2,2'-dicarbonitrile (\textbf{2})}
  \label{spectra1}
\end{figure}

\begin{figure}
 \centering
  \includegraphics[width=0.9\columnwidth]{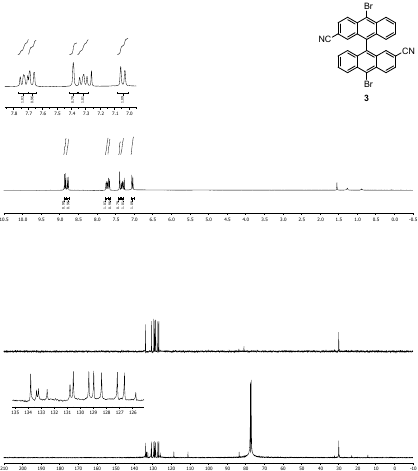}
  \caption{\textsuperscript{1}H and \textsuperscript{13}C NMR spectra of 10,10'-dibromo-[9,9'-bianthracene]-2,2'-dicarbonitrile (\textbf{3})}
  \label{spectra2}
\end{figure}

\newpage
\addcontentsline{toc}{section}{References}
\bibliography{bibliographySI}